\documentstyle [epsf]{ichep}
\begin{document}
\pagestyle{empty}

\title{
\hfill {\rm FERMILAB-Conf-94/245} \\
\bigskip
$B^+$ and $B^0$ Mean Lifetime Measurements*}

\author{Fritz DeJongh$^{\dag}$}

\affil{Fermilab\\
P.O. Box 500, Batavia IL 60510, USA}

\abstract{We review $B^+$ and $B^0$ mean lifetime measurements, including 
direct measurements and determination of the lifetime ratio via
measurements of the ratio of branching ratios.  We present world averages.}

\twocolumn[\maketitle]

\fnm{7}{Submitted to the {\it 27th Int. Conf. on High
Energy Physics}, Glasgow, Scotland, July 20-27, 1994.}
\fnm{1}{E-mail: fritzd@fnald.fnal.gov}

\section{Introduction}
The most precise determination of the CKM matrix element $V_{cb}$ 
is obtained by comparing measurements~\cite{VCB1} of the rate of
exclusive $b \rightarrow c l \nu$ decays, extrapolated to
a particular kinematic point,
with theoretical predictions~\cite{VCB2}.  The mean lifetime of the 
particular $B$ hadron for the exclusive decay is needed
to convert experimental measurements of branching ratios into decay rates.

The mean lifetime of the $D^+$ is 2.5 times that of the $D^0$.  
The difference
in the $B$ system is expected to be smaller, with the mean $B^+$ lifetime
as much as 7\% larger than that of the $B^0$~\cite{Bigi}.

We present herein results from direct measurements, as well as results
from measurements of ratios of branching ratios.  
The numerical values of all results, and the world averages, are presented
in Fig.~1.  References to a specific charge state imply the charge-conjugate
state as well.  For the direct measurements,
the decay length $L$ of the $B$ meson is measured with a silicon vertex 
detector, and the boost $\beta\gamma$ is measured with tracking in a magnetic
field, and calorimetry.  The proper lifetime $c\tau$ is simply the ratio
$L / \beta\gamma$.  We present results from CDF, from the reaction
$p\bar{p} (\sqrt{s} = 1.8~{\rm TeV}) \rightarrow b\bar{b}+X$, and results
from ALEPH, DELPHI, and OPAL, from the reaction
$e^+e^- \rightarrow Z \rightarrow b\bar{b}$.

If a $B^+$ decay mode is related to a $B^0$ decay mode by a single 
isospin amplitude, the ratio of branching ratios is equivalent to the
ratio of mean lifetimes.  
We present such results using inclusive semileptonic
decays from CLEO, and $J/\psi K$ decays from CDF.

\section{Fully reconstructed decays}
In the case where the $B$ hadron is fully reconstructed, the $B$ hadron
type, $\beta\gamma$, and $L$ are all unambiguously measured, and the
systematic uncertainties are minimal.
CDF has reconstructed~\cite{CDF1}:
\begin{itemize}
\item $B^+ \rightarrow J/\psi K^+, J/\psi K^{*+},
      \psi(2S) K^+, \psi(2S) K^{*+}$
\item $B^0 \rightarrow J/\psi K_S, J/\psi K^{*0},
\psi(2S) K_S, \psi(2S) K^{*0}$
\end{itemize}
The decay $J/\psi \rightarrow \mu^+\mu^-$ triggers the event.
Without the trigger restriction, ALEPH has reconstructed~\cite{ALEPH1}:
\begin{itemize}
\item $B^+ \rightarrow J/\psi K^+, \bar{D}^0\pi^+, 
      \bar{D}^0\rho^+,\bar{D}^0 a_1^+$
\item $B^0 \rightarrow D^- \pi^+, D^{*-}\pi^+$
\end{itemize}

\section{Partial reconstruction of $B \rightarrow \bar{D}^{(*)} l \nu X$}
In this method, one reconstructs the charmed meson
and the lepton from a semileptonic decay.
Events with $D^{*-} \rightarrow \bar{D}^0 \pi^-$ are
dominantly from $B^0$ decay, events with $\bar{D}^0$, excluding $D^{*-}$ 
candidates,
are dominantly from $B^+$ decay, and events with $D^-$ are dominantly 
from $B^0$ decay since
$\bar{D}^{*0}$ does not decay to $D^-$.

Some complications are that the decay is not fully reconstructed, so it is
necessary to estimate the boost of the $B$ based on the boost of the lepton
and charm.  This can be done with about 15\% resolution.  Systematic
uncertainties in the average boost lead to systematic uncertainties
in the mean lifetime.
Also, approximately 30\% of
semileptonic decays are through the chain $B \rightarrow \bar{D}^{**} l \nu,
\bar{D}^{**} \rightarrow \bar{D}^* \pi~{\rm or}~\bar{D} \pi$. 
This, along with an imperfect efficiency for reconstructing the soft pion 
from the $D^{*-}$
decay chain, mixes the $B^+/B^0$
composition of the samples.  This mix needs to be
understood in order to obtain individual $B$ lifetimes.

There are results from ALEPH~\cite{ALEPH2}, DELPHI~\cite{DELPHI1}, 
OPAL~\cite{OPAL}, and CDF~\cite{CDF2}.  For the purposes
of the world average, we assume that systematic uncertainties in background
shape and sample composition are correlated among all experiments.
The boost estimate depends on the $B$ production and decay spectra,
therefore we assume that systematic uncertainties from this source are 
correlated among the LEP experiments only.

\section{Topological Reconstruction}
The DELPHI collaboration has selected 1816 secondary vertex 
candidates~\cite{DELPHI2}
for which all tracks in a jet are unambiguously assigned to either
the primary or secondary vertex.  The charge of the $B$ hadron candidate
is simply the sum of the charges of the tracks assigned to the secondary 
vertex.  The $B$ purity is 99\%, and the 
charged/neutral assignment is 80\% correct for the charged sample, and
57\% correct for the neutral sample.  The proper lifetime for an event
is taken relative to the minimum value for which the event would have passed
the cuts.

The $B^+$ and $B^0$ lifetimes are extracted using independent measurements
of the $\Lambda_b$ and $B_s$ fractions and lifetimes, which lead to the
dominant systematic uncertainty.

\section{Ratios of Branching Ratios}
CLEO produces $B$ mesons in the reaction
$e^+ e^- \rightarrow \Upsilon(4S) \rightarrow B^+ B^-, B^0 \bar{B}^0$.
Therefore, tagging the charge of one $B$ determines the charge of the other
$B$, and by counting leptons opposite the tag sample, one determines the
semileptonic branching ratios of the charged and neutral $B$ mesons.
This method of using tags has the advantage over previous methods that it
does not rely on the assumption that the branching ratios
of the $\Upsilon(4S)$ to charged and neutral $B$ mesons are equal.
CLEO has used 3 tagging techniques~\cite{CLEO}:
\begin{itemize}
\item Fully reconstruct
$B \rightarrow \bar{D}^{(*)} \pi/\rho/a_1, \psi K, \psi K^*$.
\item Observe only the lepton and ``slow'' pion in the decay chain
$B^0 \rightarrow D^{*-} l \nu, D^{*-} \rightarrow \bar{D}^0 \pi^-_{\sl slow}$.
Contributions from $\bar{D}^{**}$ are suppressed with lepton momentum cuts.
This provides most of the tags for the neutral sample.
\item Reconstruct only the ``fast'' and ``slow'' pions in the decay chain
$B^0 \rightarrow D^{*-} \pi^+_{\sl fast}, 
D^{*-} \rightarrow \bar{D}^0 \pi^-_{\sl slow}$.
\end{itemize}
The major systematic uncertainties are from the uncertainty in the lepton
spectrum, and from a dependence of the efficiency of the tag selection on the
multiplicity of the untagged $B$.

CDF has reconstructed $B^+ \rightarrow \psi K^+$ and 
$B^0 \rightarrow \psi K_s$~\cite{CDF3}.
Assuming the production cross-section
for $B^+$ and $B^0$ are equal, CDF derives the ratio of branching ratios.

\section{Conclusions}
The world average is calculated grouping systematic
uncertainties into correlated sets as stated above.  The results
are shown in Fig.~1.  We conclude that the $B^+$ lifetime is consistent
with being the same as the $B^0$ lifetime.  The uncertainty on the
world average is similar to the largest theoretically expected difference.
Many experiments can add additional channels, and
all experiments cited herein continue to take data, so we can expect 
continued improvements in precision in the near future.

\section{Acknowledgements}
Thanks to Kay Kinoshita, Robert Kowalewski, Vivek Sharma, 
Fumi Ukegawa, and Wilbur Venus for providing 
information on the latest results.
Thanks to Tim Hessing for the use of a program to calculate averages
with correlated systematic uncertainties.

\newpage
\begin{figure*}
\epsffile[50 50 700 700]{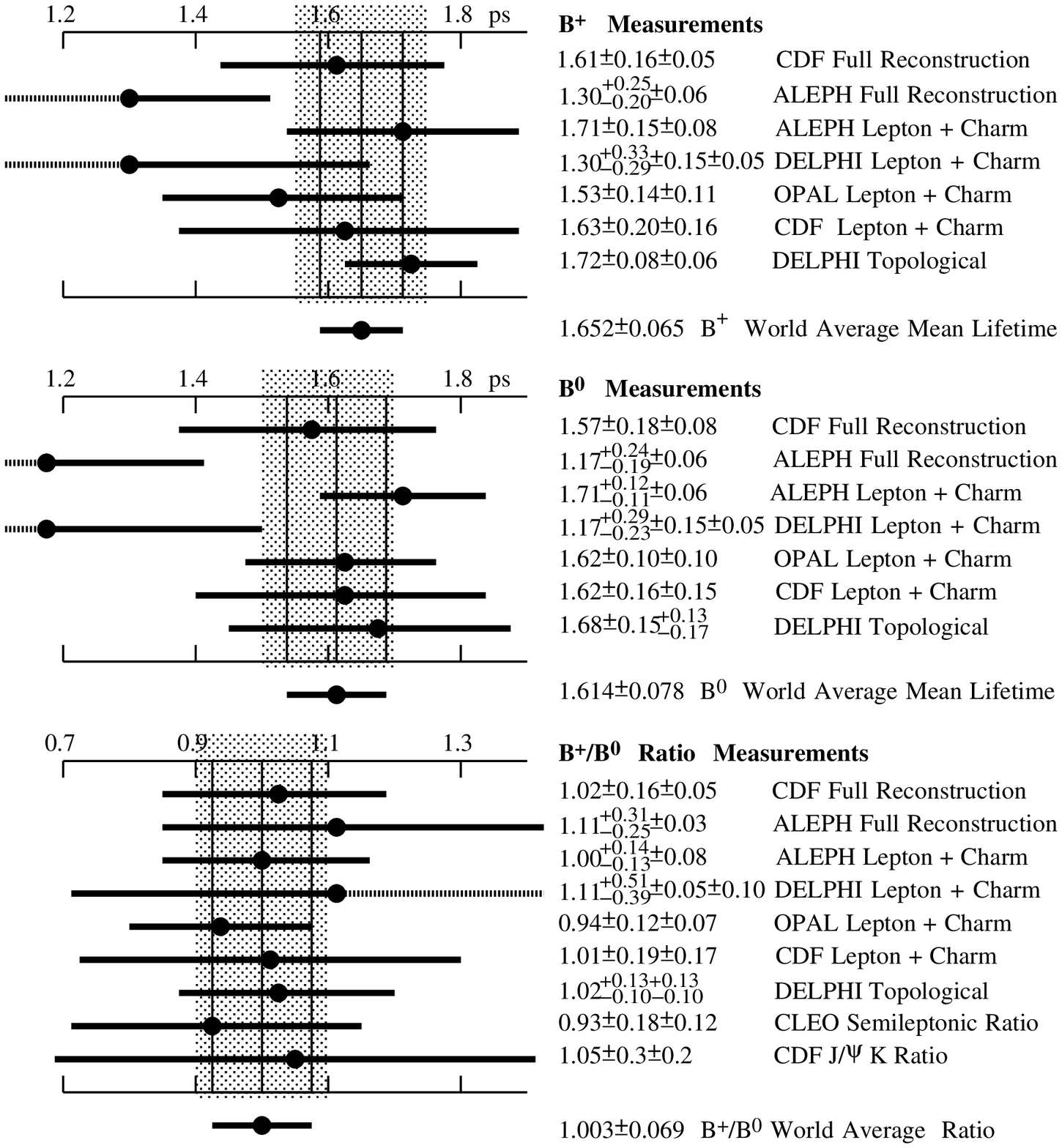}
\caption{{\label{fig:fdjctau1}}
Summary of $B^+$ and $B^0$ mean lifetime measurements.}
\end{figure*}
\clearpage


\Bibliography{9}
\bibitem{VCB1} CLEO collaboration, contribution to this conference,
 ICHEP94 Ref. 0251;  ALEPH collaboration, 
 contribution to this conference, ICHEP94 Ref. 0605.
\bibitem{VCB2}
 M. Shifman, contribution to this conference, ICHEP94 Ref. 0823;
 M. Neubert, contribution to this conference, ICHEP94 Ref. 0935.
\bibitem{Bigi} I.I. Bigi and N.G. Uraltsev, \pl{B280}{92}{271}.
\bibitem{CDF1} F. Abe {\it et al.}, \prl{71}{94}{3456}.
\bibitem{ALEPH1} ALEPH collaboration, 
 contribution to this conference, ICHEP94 Ref. 0602.
\bibitem{ALEPH2} ALEPH collaboration,
 contribution to this conference, ICHEP94 Ref. 0579.
\bibitem{DELPHI1} P. Abreu {\it et al.}, \zp{C57}{93}{181}.
\bibitem{OPAL} OPAL collaboration, 
 contribution to this conference, ICHEP94 Ref. 0533.
\bibitem{CDF2} CDF collaboration, 
 contribution to this conference, ICHEP94 Ref. 0655.
\bibitem{DELPHI2} DELPHI collaboration, 
 contribution to this conference, ICHEP94 Ref. 0165.
\bibitem{CLEO} M. Athanas {\it et al.}, CLEO 94-16.
\bibitem{CDF3} J. Gonzalez, Proc. of the Workshop on $B$ Phys. at Hadron
 Accelerators 1993, p. 285; Eds. P. McBride and C.S. Mishra.
\end{thebibliography}
\end{document}